\documentclass[11pt]{article}
\usepackage{amssymb}
\usepackage{epsfig}

\newcommand{\BE}{\begin{equation}}
\newcommand{\EE}{\end{equation}}
\newcommand{\BA}{\begin{eqnarray}}
\newcommand{\EA}{\end{eqnarray}}

\begin{document}
\begin{titlepage}

\vspace*{1mm}
\begin{center}

            {\LARGE{\bf From classical to modern ether-drift experiments: \\
the narrow window for a preferred frame }}

\vspace*{14mm}
{\Large  M. Consoli and E. Costanzo}
\vspace*{4mm}\\
{\large
Istituto Nazionale di Fisica Nucleare, Sezione di Catania \\
Dipartimento di Fisica e Astronomia dell' Universit\`a di Catania \\
Via Santa Sofia 64, 95123 Catania, Italy}
\end{center}
\begin{center}
{\bf Abstract}
\end{center}
Modern ether-drift experiments look for a preferred frame by
measuring the difference $\Delta \nu$
in the relative frequencies of two cavity-stabilized lasers, upon local
rotations of the apparatus or under the Earth's rotation. 
If the small deviations observed in the classical
ether-drift experiments were not mere instrumental artifacts, by
replacing the high vacuum in the resonating cavities 
with a dielectric gaseous medium (e.g. air), the typical measured 
$\Delta\nu\sim 1$ Hz should increase {\it by orders of magnitude}. This 
prediction is consistent with the characteristic modulation of a few kHz
observed in the original experiment with He-Ne masers. 
However, if such enhancement would not be confirmed by new and more
precise data, the existence of a preferred frame can be definitely ruled out.
 
\vskip 35 pt
PACS: 03.30.+p, 01.55.+b
\vskip 140 pt
To appear in Physics Letters A
\end{titlepage}


{\bf 1.}~The controversy about the existence of a preferred reference frame 
dates back to the birth of the Theory of Relativity, i.e. to the basic
differences between Einstein's Special Relativity \cite{einstein}
and the Lorentz-Poincar\'e point of view \cite{lorentz,poincare}. 
Today the former interpretation is generally
accepted. However, the conceptual relevance of retaining
a physical substratum as an important element of the physical theory 
\cite{martin}, may induce to re-discover the potentially
profound implications of the latter \cite{bell,brown}.
For instance, replacing the empty space-time of
Special Relativity with a preferred frame, one gets
a different view of the non local aspects of the quantum theory, 
see Refs.\cite{hardy,scarani}.

Another argument that might induce to re-consider
the idea of a preferred frame was given in Ref.\cite{pagano}. The argument
was based on the simultaneous presence of two ingredients that are often
found in present-day elementary particle physics, namely: a) vacuum 
condensation, as with the Higgs field in the electroweak theory, and b) an
approximate form of locality, as with cutoff-dependent, effective quantum
field theories. In this case, one is faced with `reentrant violations of
special relativity in the low-energy corner' \cite{volo}. These are deviations
at {\it small} momenta $|{\bf{p}}| < \delta$ where the infrared scale $\delta$
vanishes, in units of the Lorentz-invariant scale $M$ of the theory, 
only in the local limit of the continuum theory
${{\Lambda}\over{M}} \to \infty$, $\Lambda$ being the
ultraviolet cutoff. A simple interpretation of the phenomenon, 
 in the case of a condensate
of spinless quanta, is in terms of density fluctuations of the 
system \cite{stevenson,consoli}, the continuum theory corresponding to the
incompressibility limit. The resulting
picture of the ground state is closer to a medium with a non-trivial
refractive index \cite{pagano} than to the empty space-time of
Special Relativity. 

Therefore, in the presence of a non-trivial vacuum, it is natural to explore
whether the {\it physically realized} form of the Theory of
Relativity is closer to the Einstein's formulation or to the original Lorentzian
approach with a preferred frame. In other words, 
the same relativistic effects between two observers
$S'$ and $S''$, rather than being due to their {\it relative} motion, might
be interpreted as arising from
their {\it individual} motion with respect to some preferred
frame $\Sigma$. This equivalence
is a simple consequence of the group structure of Lorentz transformations, 
where the relative velocity parameter $\beta_{\rm rel}$
connecting $S'$ to $S''$ can be expressed in terms of the
individual velocity parameters $\beta'$ and $\beta''$ respectively
relating $S'$ and $S''$ to $\Sigma$ as
\BE
          \beta_{\rm rel}= {{\beta' - \beta''}\over{ 1- \beta' \beta''}}
\EE
(we restrict for simplicity to one-dimensional motions). 

In this perspective, 
 the crucial question becomes the following: can the individual parameters 
$\beta'$ and $\beta''$ be determined separately through ether-drift 
experiments ? 
Accepting the standard `null-result' interpretation of this type of
experiments, this is not possible. 
Therefore, if really only $\beta_{\rm rel}$ is experimentally measurable, 
one is driven to conclude (as Einstein did in 1905 
\cite{einstein}) that the introduction of a preferred frame is `superfluous',
all effects of $\Sigma$ being re-absorbed into the relative space-time
units of any pair $(S',S'')$. 

On the other hand, if the ether-drift experiments give a non-null
result, so that $\beta'$ and $\beta''$ can be separately determined, then
the situation is completely
different. In fact, now $\beta_{\rm rel}$ is a derived quantity
and the Lorentzian point of view is uniquely singled out. 

Due to the importance of the problem, we have first
re-considered the classical ether-drift experiments, 
our main motivation being that, 
according to some authors, their standard null-result 
interpretation is not so obvious. 
The fringe shifts observed in the various Michelson-Morley type of
experiments, although smaller than the classical predictions,
were never really negligibly small. Interpreting these small deviations
on the base of
Ref.\cite{pagano}, a narrow experimental window might still be compatible
with the existence of a preferred frame.

After this first part, we have concentrated our analysis on 
the modern ether-drift experiments, those 
where the observation of the interference
fringes is replaced by the difference $\Delta \nu$
in the relative frequencies of two cavity-stabilized lasers upon local
rotations of
the apparatus \cite{brillet} or under the Earth's rotation \cite{muller}. It
turns out that, even in this case, 
the most recent data \cite{muller} leave some space for a non-null 
interpretation of the experimental results. 

For this reason, we shall  
propose a sharp experimental test that can definitively decide about
the existence of a preferred frame. If the small deviations found in the
classical experiments were not mere instrumental artifacts, 
by replacing the high vacuum used in the 
resonating cavities with a dielectric gaseous medium, the typical
frequency of the signal should increase 
from values $\Delta\nu\sim 1$ Hz up to $\Delta\nu \sim 100$ kHz, using air, or
up to $\Delta\nu \sim 10$ kHz, using helium. The latter 
prediction appears to be consistent with the characteristic modulation of a 
few kHz in the magnitude of the $\Delta\nu$'s observed by Jaseja et al. \cite{jaseja}
using He-Ne masers. 
\vskip 15 pt

{\bf 2.}~A non-null result of the original Michelson-Morley \cite{mm}
experiment was strongly advocated by Hicks \cite{hicks} long time ago.
The same conclusion was obtained by Miller after his re-analysis of
the Michelson-Morley data, of the Morley-Miller \cite{morley} experiments 
and of his own measurements at Mt.Wilson, 
see Fig.4 of Ref.\cite{miller}. Miller's refined analysis 
showed that all data were consistent with an
effective, {\it observable} velocity lying in the range 7-10 km/s, say
\BE
\label{vobs}
       v_{\rm obs} \sim 8.5 \pm 1.5~{\rm km/s}
\EE
For comparison, the Michelson-Morley experiment gave a value
$v_{\rm obs} \sim 8.8 $ km/s for the noon observations 
and a value $v_{\rm obs} \sim 8.0 $ km/s for the evening observations. 
As the
fringe shifts grow quadratically with the velocity, their typical magnitude was 
$(8.5/30)^2\sim$ 1/13 of that expected, on the base of
classical physics, for the Earth's orbital velocity of 30 km/s.

The difference of the value in Eq.(\ref{vobs}) with
       respect to the original conclusion of Michelson-Morley
($v_{\rm obs}$ certainly smaller 
than 1/4 of the Earth's orbital velocity \cite{mm}), can partly be
understood looking at the conclusions of the
 Hicks' study \cite{hicks}:
     one is not allowed to average data of different experimental 
     sessions unless 
     one is sure that the direction of the ether-drift effect remains the same
     (see page 34 of \cite{hicks} ``It follows that averaging the results of
     different days in the usual manner is not allowable...If this is not 
     attended to, the average displacement may be expected to come out 
     zero...''). In other words, the ether-drift, if it exists, has
a vectorial nature. Therefore, rather than averaging
the raw data from the various sessions, one should first consider
the data from the i-th experimental session and extract the observable velocity 
$v_{\rm obs}(i)$ and the ether-drift direction $\theta_o(i)$ 
for that session. Finally, a mean magnitude 
$\langle v_{\rm obs}\rangle$ and a mean direction $\langle \theta_o \rangle$
can be obtained by averaging the individual determinations
(see Figs. 22 of Ref.\cite{miller}).

Now, following the latter strategy, 
       the magnitude of the observable velocity 
        comes out to be larger, its error becomes 
       smaller so that the evidence for an ether-drift effect becomes stronger
(see page 36 of  Ref.\cite{hicks} `` ...this naturally leads 
     to the reconsideration of the numerical data obtained by Michelson 
     and Morley, who did lump together the observations taken in 
     different days. I propose to show that, instead of giving a null 
     result, the numerical data published in their paper show distinct 
     evidence of an effect of the kind to be expected''). 

The same was true for the Morley-Miller data \cite{morley}. In this case, 
the morning and evening observations each were indicating
an effective velocity of
about 7.5 km/s (see Fig.11 of Ref.\cite{miller}). This indication
was completely lost after 
averaging the raw data as in Ref.\cite{morley}. Finally, 
the same point of view has been advocated by M\'unera in his recent 
re-analysis of the classical experiments \cite{munera}. 

\vskip 15 pt

{\bf 3.}~
Now, suppose we accept
the value in Eq.(\ref{vobs}) to summarize the results of the
Michelson-Morley, Morley-Miller and Miller experiments. 
As these were performed in air, it would mean that 
the measured 
two-way speed of light differs from an exactly isotropical value
\BE
\label{air}
u_{\rm air}={{c}\over{\cal{N}_{\rm air} }}
\EE
${\cal N}_{\rm air}$ denoting the refractive index of the air. Namely, 
for an observer placed on the Earth (where the air is at rest or more 
precisely in thermodynamical equilibrium) the experiments say that there is
a small anisotropy at the level
${\cal O}({{v^2_{\rm obs}}\over{c^2}}) \sim 10^{-9}$ so that the isotropical
value Eq.(\ref{air}) is only accurate at a lower level of accuracy, say
$\sim 10^{-8}$. 

On the other hand, for the Kennedy's \cite{kennedy0} experiment, 
where the whole
optical system was inclosed in a sealed metal case containing helium at 
atmospheric pressure, the observed anisotropy was definitely smaller. In fact, 
the accuracy of the experiment, such to exclude
 fringe shifts as large as 1/4 of those expected 
on the base of Eq.(\ref{vobs}) (or 1/50 of that expected on the base of a
velocity of 30 km/s) allows to place an upper bound
$v_{\rm obs} < 4$ km/s. This is confirmed by the re-analysis of the
Illingworth's experiment \cite{illing} performed by M\`unera \cite{munera}
who pointed out some incorrect assumptions in the original analysis of the 
data. From this re-analysis, the relevant observable velocity turns out to be 
$v_{\rm obs}= 3.1 \pm 1.0$ km/s (errors at the 95$\%$ C.L.) 
\cite{munera}, with typical fringe shifts that were 1/100 of that expected
for a velocity of 30 km/s.
Again, this means that, 
for an apparatus filled with gaseous helium at atmospheric pressure, 
the measured two-way speed of light differs from the exactly isotropical value
${{c}\over{\cal{N}_{\rm helium} }}$ by terms
${\cal O}({{v^2_{\rm obs}}\over{c^2}}) \sim 10^{-10}$. 

Finally, for the Joos experiment \cite{joos}, performed in an evacuated
housing and where any ether-wind was found smaller than $1.5$ km/s, 
the typical value $v_{\rm obs}\sim 1$ km/s means
that, in that particular type of vacuum, the fringe shifts were smaller than
1/400 of those expected for an Earth's velocity of 30 km/s and
the anisotropy of the two-way speed of light was at the 
level $\sim 10^{-11}$. 

We shall try to summarize the above experimental 
results as follows. When light propagates in 
a gaseous medium, the exactly isotropical value
\BE
\label{u}
u={{c}\over{\cal{N}_{\rm medium} }}
\EE
holds approximately for an observer placed on the Earth. Within the context of
a Theory of Relativity with a preferred frame, 
this should not come as a complete surprise. In fact, 
the usual assumption, that the isotropical value
Eq.(\ref{u}) holds exactly in the reference frame $S'$
where the gas is in thermodynamical equilibrium, reflects the
point of view of Special Relativity with no preferred frame. However, 
to {\it test} this assumption requires precisely to
perform a Michelson-Morley experiment and look 
for fringe shifts upon local rotations of the apparatus. 

When this is done, the experiments indicate a slight anisotropy that 
becomes smaller when the refractive index of the medium 
approaches unity. In fact $v_{\rm obs}$, and thus the anisotropy,
is larger for those interferometers operating in air, where
${\cal N}_{\rm air} \sim 1.00029$, and becomes smaller in 
experiments performed in helium, where
${\cal N}_{\rm helium}\sim 1.000036$, or in an evacuated housing.
This is completely consistent with the expectations based on 
Lorentz transformations that preserve the isotropical 
value of the speed of light 
in the vacuum $c=2.9979...\cdot10^{10}$ cm/s. 
If these are valid, even in the presence of a preferred frame, 
no anisotropy can be detected studying light propagation in the vacuum 
where ${\cal N}_{\rm vacuum}=1$ identically.

However, Lorentz transformations 
do not preserve the value of the speed of light in a medium.
Therefore, the simplest way to generate an anisotropy in $S'$ is  
to start from the isotropical value Eq.(\ref{u}), assumed to be valid
in some preferred frame $\Sigma$, and 
compute its value in $S'$ through a Lorentz transformation. 

Notice that this series of steps is completely analogous to 
the conventional treatment of the Michelson-Morley experiment. There, 
one starts from the isotropical value $c$ in 
$\Sigma$ and uses Galileian relativity
(for which the speed of light becomes $c\pm v$)
to transform to the observer $S'$ placed in the Earth's frame. 
Here we shall only take into account that
i) light propagates in 
a gaseous medium and ii) Galilei's trasformations have to be
replaced by Lorentz transformations. 

There is, however, a hidden assumption in our procedure
that should be clearly spelled out. Eq.(\ref{u}) is strictly valid for a 
medium at rest in the preferred frame $\Sigma$. However, 
the medium is at rest in $S'$ and not in $\Sigma$. 
Therefore, strictly speaking, before Lorentz transforming to $S'$, 
we should first correct the $\Sigma$ estimate for the effect of the Fresnel's
drag that might exist anyway. 
Of course, if we had to use the exact relativistic 
formula to compute this effect and then transform to
$S'$, we would obtain that the isotropical value 
Eq.(\ref{u}) holds in $S'$ as well. 
Here, following the experimental indications of a non-zero anisotropy in
$S'$, we shall assume that the Fresnel's drag for $\Sigma$ 
is negligible, at least for gaseous media, so that the 
anisotropy in $S'$ is due to the genuine 
Lorentz transformation. This assumption reflects the point of view 
that, if there is really a preferred frame, there must be somewhere 
a basic asymmetry between $\Sigma$ and $S'$. We stress, however, that  
our assumption of a negligible Fresnel's
drag in $\Sigma$ cannot be extended to light propagation in 
solid dielectrics with high refractive index. In fact,
Michelson-Morley experiments performed in a solid transparent medium
(perspex) \cite{fox}, where
${\cal N}_{\rm perspex}\sim 1.5$, show no anisotropy. 

 The peculiar role of gaseous media can partly be understood noticing that they
 cover the ideal limit where the refractive index ${\cal N}$ tends
 to unity. For ${\cal N}=1$, light is seen to propagate 
 isotropically in the hypothetical preferred frame and in all moving frames
 $S'$, $S''$, $S'''$,...For ${\cal N}\neq 1$, however, light 
 has to resolve this infinite `degeneracy' and
 necessarily choose between two different alternatives: either to 
 propagate isotropically in the rest frame of the medium or in 
 $\Sigma$. The experiments suggest that for $({\cal N} - 1) \ll 1$
 there might be still no Fresnel's drag for $\Sigma$ so that
 light is seen to propagate isotropically in $\Sigma$ and not in $S'$. 
 However, when ${\cal N}$ starts to differ sizeably from unity, the Fresnel's
 drag in $\Sigma$ becomes substantial so as to cancel the effect of the genuine
 Lorentz transformation to $S'$. 
 In this new regime, light propagates isotropically in
 the rest frame of the medium. Just for this reason, 
 experiments performed in gaseous media represent the only remaining 
 window to detect the possible existence of a preferred frame. 

Within the above assumptions, starting from Eq.(\ref{u}) and
denoting by ${\bf{v}}$ the velocity of
$S'$ with respect to $\Sigma$, a Lorentz transformation give 
the general expression for the one-way speed of light in $S'$ 
($\gamma= 1/\sqrt{ 1- {{ {\bf{v}}^2}\over{c^2}} }$) 
\BE
\label{uprime}
  {\bf{u}}'= {{  {\bf{u}} - \gamma {\bf{v}} + {\bf{v}}
(\gamma -1) {{ {\bf{v}}\cdot {\bf{u}} }\over{v^2}} }\over{ 
\gamma (1- {{ {\bf{v}}\cdot {\bf{u}} }\over{c^2}} ) }}
\EE
where $v=|{\bf{v}}|$. By keeping terms up 
to second order in $v/u$, denoting by
$\theta$ the angle between ${\bf{v}}$ and ${\bf{u}}$
and defining $u'(\theta)= |{\bf{u'}}|$, 
we obtain
\BE
  {{ u'(\theta) }\over{u}}= 1- \alpha {{v}\over{u}} -\beta {{v^2}\over{u^2}}
\EE
where 
\BE
   \alpha = k_{\rm medium} \cos \theta + 
{\cal O} (k^2_{\rm medium} )
\EE
\BE
\beta = k_{\rm medium} P_2(\cos \theta) +
{\cal O} ( k^2_{\rm medium} )
\EE
with 
\BE
 k_{\rm medium}=
1- {{1}\over{ {\cal N}^2_{\rm medium} }} \ll 1
\EE
and $P_2(\cos \theta) = {{1}\over{2}} (3 \cos^2\theta -1)$. 

Finally, the two-way speed of light is 
\BE
\label{twoway}
{{\bar{u}'(\theta)}\over{u}}= {{1}\over{u}}~ {{ 2  u'(\theta) u'(\pi + \theta) }\over{ 
u'(\theta) + u'(\pi + \theta) }}= 1- {{v^2}\over{c^2}} ( A + B \sin^2\theta) 
\EE
where 
\BE 
\label{ath}
   A= k_{\rm medium}  + {\cal O} ( k^2_{\rm medium} )
\EE
and 
\BE
\label{BTH}
     B= -{{3}\over{2}} k_{\rm medium} 
+ {\cal O} (k^2_{\rm medium} )
\EE
In this way, as shown in Ref.\cite{pagano}, one obtains formally the same
pre-relativistic expressions where the
kinematical velocity $v$ is replaced by an effective observable velocity
\BE
\label{vobs0}
           v_{\rm obs}= v
\sqrt { k_{\rm medium} }
 \sqrt{3} \sim v \sqrt{ -2B}
\EE
For instance, 
for the Michelson-Morley experiment, and for an ether wind
along the $x$ axis, 
the prediction for the fringe shifts at a given angle 
$\theta$ with the $x$ axis
has the particularly simple form ($D$ being the length of each arm
of the interferometer as measured in $S'$) 
\BE 
\label{fringe}
{{\Delta \lambda (\theta)}\over{\lambda}}=
{{u}\over{\lambda}} (
{{2D }\over{\bar{u'}(\theta)}}- {{2 D }\over{\bar {u'}(\pi/2+\theta)}})\sim 
 {{ D }\over{\lambda}} {{v^2}\over{c^2}} (-2B) \cos (2\theta) =
 {{ D }\over{\lambda}} {{v^2_{\rm obs}}\over{c^2}}\cos (2\theta) 
\EE
that corresponds to a pure second-harmonic 
effect as in the old theory (see for instance 
Refs.\cite{hicks,kennedy}) where $v^2$ is replaced by $v^2_{\rm obs}$. 
Notice that, in agreement with the basic isotropy of space, 
embodied in the validity of Lorentz transformations, the measured
length of an interferometer at rest in $S'$ is $D$ regardless of its 
orientation. 

Also, the trend predicted by 
Eqs.(\ref{vobs0}) and (\ref{fringe}), where
the observable velocity, and thus the anisotropy, becomes smaller
and smaller when ${\cal N}_{\rm medium}$ approaches unity, 
is consistent with the analysis of the experiments performed by 
Kennedy, Illingworth and Joos vs. those of
Michelson-Morley, Morley-Miller and Miller.
We note that a qualitatively 
similar suppression effect had already been discovered by
Cahill and Kitto \cite{cahill} by following a different approach. 

\vskip 15 pt

{\bf 4.}~As stressed in Ref.\cite{pagano}, the detection of a 
preferred frame in ether-drift experiments 
is a purely experimental issue. Within our assumptions, 
this requires: i) the preliminary
observation of fringe shifts
upon operation of the interferometer and ii) that
their magnitude, observed with different gaseous media and within the
experimental errors, points
consistently to a unique value of the kinematical Earth's velocity. Only
in this case, one can conclude that there is
experimental evidence for
the existence of a preferred frame. 

To extract the value of the kinematical Earth's velocity corresponding to
the various $v_{\rm obs}$, one should
re-analyze the experiments in terms of the effective parameter
 $\epsilon={{v^2_{\rm earth} }\over{u^2}} k_{\rm medium}$. 
The conclusion of Cahill and Kitto \cite{cahill} 
is that the classical experiments are consistent with the value
 $v_{\rm earth}\sim 365$ km/s obtained from the dipole fit to the
COBE data \cite{cobe} for the cosmic background radiation.
However, in
our expression Eq.(\ref{vobs0}) determining the fringe shifts there is a 
difference
of a factor $\sqrt{3}$ with respect to their result
$v_{\rm obs}=v \sqrt { k_{\rm medium} }$. Therefore, using
Eqs.(\ref{vobs0}) and (\ref{vobs}), for 
${\cal N}_{\rm air} \sim 1.00029$, 
 the relevant Earth's velocity (in the plane of the interferometer)
 is {\it not} $v_{\rm earth}\sim 365$ km/s but rather
\BE
\label{vearth}
                  v_{\rm earth} \sim 204 \pm 36 ~{\rm km/s}
\EE
In this way, using our Eq.(\ref{vobs0}), 
the kinematical Earth's velocity becomes consistent with the 
values needed by Miller to understand the {\it variations} 
of the ether-drift effect in different epochs of the year \cite{miller}.
In fact, the typical
daily values, in the plane of the interferometer, had to lie in the range
$ 195 \le v_{\rm earth} \le 211$ {\rm km/s} 
(see Table V of Ref.\cite{miller}).  
Such a consistency, on one hand, 
increases the body of experimental evidence for a preferred frame, and
on the other hand, provides a definite range of velocities 
to be used in the analysis of the other experiments. 

To this end, let us compare with the experiment performed by Michelson, 
Pease and Pearson \cite{mpp}. These other authors in 1929, 
using their own interferometer, again at Mt. Wilson, declared that 
their ``precautions taken to eliminate effects of 
temperature and flexure disturbances were effective''. Therefore, their statement that the
fringe shift, 
as derived from ``...the displacements observed at maximum and minimum at 
sidereal times...'', was definitely smaller than ``...one-fifteenth of
that expected on the 
supposition of an effect due to a motion of the Solar System of three 
hundred kilometres per second'', can be taken as an indirect 
confirmation of our Eq.(\ref{vearth}). Indeed,
although the ``one-fifteenth'' was actually 
a ``one-fiftieth'' (see page 240 of Ref.\cite{miller}), 
their fringe shifts were certainly non negligible. This is easily understood 
since, for an in-air-operating interferometer, 
the fringe shift $(\Delta\lambda)_{\rm class}(300)$, expected on the base of 
classical physics
for an Earth's velocity of 300 km/s, is about 500 times
bigger than the corresponding relativistic one
\BE
(\Delta\lambda)_{\rm rel}(300)\equiv 3 k_{\rm air}
~ (\Delta\lambda)_{\rm class}(300)
\EE
computed using Lorentz transformations 
(compare with Eq.(\ref{fringe}) for  
$k_{\rm air}\sim {\cal N}^2_{\rm air} -1 \sim 0.00058$). 
 Therefore, the Michelson-Pease-Pearson upper bound
\BE
(\Delta\lambda)_{\rm obs}< 0.02~
 (\Delta\lambda)_{\rm class} (300)
\EE
is actually equivalent to
\BE
(\Delta\lambda)_{\rm obs}< 24 ~
 (\Delta\lambda)_{\rm rel} (204)
\EE
As such, it poses no strong restrictions and is entirely 
consistent with those typical low observable velocities reported in 
Eq.(\ref{vobs}).

A similar agreement is obtained when comparing with the Illingworth's data
\cite{illing} as recently re-analyzed by M\'unera \cite{munera}. 
In this case, using Eq.(\ref{vobs0}), 
the observable velocity $v_{\rm obs}=3.1 \pm 1.0$ km/s \cite{munera}
(errors at the 95$\%$ C.L.)  
and the value ${\cal N}_{\rm helium}-1 \sim 3.6\cdot 10^{-5}$, 
one deduces $v_{\rm earth}=213 \pm 36$ km/s
(errors at the 68$\%$ C.L.) in very good agreement with our
Eq.(\ref{vearth}). 

The same conclusion applies to the Joos experiment 
\cite{joos}. Although we don't know the exact 
value of ${\cal N}_{\rm vacuum}$ for the Joos experiment, 
it is clear that his result, $v_{\rm obs}<$ 1.5 km/s, represents
the natural type of upper bound
in this case. As an example, for $v_{\rm earth}\sim 204$ km/s, one
obtains $v_{\rm obs}\sim 1.5$ km/s for 
${\cal N}_{\rm vacuum}-1= 9\cdot 10^{-6}$ and 
$v_{\rm obs}\sim 0.5$ km/s for 
${\cal N}_{\rm vacuum}- 1=1\cdot 10^{-6}$. 
In this sense, the effect of using Lorentz 
transformations is most 
dramatic for the Joos experiment when comparing with
the classical expectation for an Earth's velocity
of 30 km/s. Although the relevant Earth's velocity 
can be as large as $204$ km/s, 
the fringe shifts, rather than being $(204/30)^2\sim 50$ times {\it bigger} than the 
classical prediction, are  
$\sim (30/1.5)^2= 400$ times {\it smaller}. 

\vskip 15 pt
{\bf 5.}~Let us finally consider those present-day, 
 `high vacuum' Michelson-Morley experiments of the type first performed by 
Brillet and Hall \cite{brillet} and more recently by 
M\"uller et al. \cite{muller}. 
In these experiments, the test of the isotropy 
of the speed of light does not consist in the observation of the interference
fringes as in the classical experiments. 
Rather, one looks for the difference
$\Delta \nu$
in the relative frequencies of two cavity-stabilized lasers upon local
rotations of
the apparatus \cite{brillet} or under the Earth's rotation \cite{muller}. 

The present experimental value for the anisotropy of the two-way speed of light
in the vacuum, as determined by M\"uller et al.\cite{muller}, 
\BE
\label{experiment}
{{\Delta \nu}\over{\nu}}=
        ({{ \Delta \bar{c}_\theta }\over{c}})_{\rm exp}= (2.6 \pm 1.7) \cdot 10^{-15}
\EE
can be interpreted
within the framework of our Eq.(\ref{twoway}) where
\BE
        ({{ \Delta \bar{c}_\theta }\over{c}})_{\rm theor} \sim 
|B_{\rm vacuum}| {{v^2_{\rm earth} }\over{c^2}} 
\EE
Now, in a perfect vacuum by definition 
${\cal N}_{\rm vacuum}=1$ so that
$B_{\rm vacuum}$ and  $v_{\rm obs}$ vanish. 
However, one can explore 
\cite{pagano} the possibility that, even in this case,
 a very small anisotropy might be due to a refractive index 
${\cal N}_{\rm vacuum}$ that differs from unity by an infinitesimal
amount. In this case, the natural candidate to explain a value
${\cal N}_{\rm vacuum} \neq 1$
is gravity. In fact, by using
the Equivalence Principle, a freely falling frame $S'$ will locally 
measure the same speed of light as in an inertial frame in the absence of
any gravitational effects. However, if $S'$ carries on board an heavy 
object this is no longer true. For an observer placed on the Earth, 
this amounts to insert
the Earth's gravitational potential in the  weak-field isotropic
approximation to the line element of
 General Relativity \cite{weinberg}
\BE
ds^2= (1+ 2\varphi) dt^2 - (1-2\varphi)(dx^2 +dy^2 +dz^2)
\EE
so that one obtains a refractive index for
light propagation 
\BE
\label{nphi}
            {\cal N}_{\rm vacuum}\sim  1- 2\varphi
\EE
This represents the `vacuum analogue' of 
${\cal N}_{\rm air}$, ${\cal N}_{\rm helium}$,...so that from
\BE 
     \varphi =- {{G_N M_{\rm earth}}\over{c^2 R_{\rm earth} }} \sim
-0.7\cdot 10^{-9}
\EE
and using Eq.(\ref{BTH}) one predicts
\BE
\label{theor}
                 B_{\rm vacuum} \sim -4.2 \cdot 10^{-9}
\EE
Adopting the range
of Earth's velocity (in the plane
of the interferometer) given in Eq.(\ref{vearth}) this leads to predict
an observable anisotropy of the two-way speed of light 
in the vacuum Eq.(\ref{twoway}) 
\BE
        ({{ \Delta \bar{c}_\theta }\over{c}})_{\rm theor} \sim 
|B_{\rm vacuum}| {{v^2_{\rm earth} }\over{c^2}} \sim (1.9\pm 0.7)\cdot 10^{-15}
\EE
consistently with the experimental value in Eq.(\ref{experiment}). 

Clearly, in this framework, trying to rule out the existence of
a preferred frame through the experimental 
determination of ${{ \Delta \bar{c}_\theta }\over{c}}$ in a high
vacuum is not the most convenient strategy 
due to the vanishingly small value of
$B_{\rm vacuum}$. For this reason, a more efficient search 
might be performed in dielectric gaseous media.
As a check, we have compared with the only available results
obtained by Jaseja et. al \cite{jaseja} in 1963 when
looking at the relative frequency shifts of two orthogonal 
He-Ne masers placed on a rotating platform. 
As we shall show in the following, their data are consistent with the same type
of conclusion obtained from the classical experiments: an ether-drift effect 
determined by an Earth's velocity as in Eq.(\ref{vearth}). 

To use the experimental results reported by Jaseja et al.\cite{jaseja}
one has to subtract preliminarly a large overall systematic 
effect that was present in their data and interpreted by the authors as
probably due to magnetostriction
in the Invar spacers induced by the Earth's magnetic field. 
As suggested by the same authors, this spurious
effect, that was only affecting the normalization of the experimental
$\Delta \nu$, can be subtracted looking at the variations of the data
at different hours of the day. The data for $\Delta\nu$,
in fact, in spite of their rather large errors,
exhibit a characteristic modulation (see Fig.3 of Ref.\cite{jaseja})
with a maximum at about 7:30 a.m. and a minimum at about
9:00 a.m. and a typical difference \cite{jaseja}
\BE
\label{resultb}
 \delta(\Delta\nu)\sim (1.6\pm 1.2)~ {\rm kHz}
\EE
Our theoretical starting point to understand the above 
(rather loose) determination
is the formula for the frequency shift of the two
masers at an angle $\theta$ with the direction of the ether-drift 
\BE
\label{prediction}
              {{\Delta \nu (\theta) }\over{\nu}}= 
{{\bar{u}'(\pi/2 +\theta)- \bar{u}' (\theta)} \over{u}}=
|B_{\rm He-Ne}| {{v^2_{\rm earth} }\over{c^2}} \cos(2\theta)
\EE
where, taking into account the values
${\cal N}_{\rm helium}\sim 1.000036$, ${\cal N}_{\rm neon}\sim 1.000067$, 
${\cal N}_{\rm He-Ne}\sim 1.00004$ and Eq.(\ref{BTH}) we shall use
$|B_{\rm He-Ne}|\sim 1.2\cdot 10^{-4}$. 

Further, using the value of the frequency of Ref.\cite{jaseja} 
$\nu\sim 3\cdot 10^{14}$ Hz and our
standard value Eq.(\ref{vearth})
for the Earth's velocity in the plane of the interferometer 
$v_{\rm earth} \sim 200$ km/s, Eq.(\ref{prediction}) leads to
the reference value for the amplitude of the signal 
\BE
\label{reference}
(\Delta \nu)_{\rm ref}= \nu
|B_{\rm He-Ne}| {{(200~ {\rm km/s})^2 }\over{c^2}} \sim 16 ~{\rm kHz}
\EE
and to its time modulation
\BE
\label{prediction2}
 \delta(\Delta\nu)_{\rm theor}
 \sim 16~{\rm kHz} ~{{\delta v^2}\over{v^2}}
\EE
where
\BE
\label{deltav}
{{\delta v^2}\over{v^2}}\equiv 
 {{v^2_{\rm earth} (7:30~{\rm a.m.}) -v^2_{\rm earth} (9:00~{\rm a.m.}) }
\over{(200~{\rm km/s})^2}}
\EE
To evaluate the above ratio of velocities, let us first 
compare the modulation of $\Delta\nu$ seen in fig.3 of ref.\cite{jaseja} 
with that of $v_{\rm obs}$ in
fig.27 of ref.\cite{miller} (data plotted as a function of civil time as
in ref.\cite{jaseja}) restricting to the 
Miller's data of February, the period of the year that is
closer to the date of January 20th when Jaseja et {\it al.} performed their
experiment. Further, 
the different location of the two laboratories (Mt.Wilson
and Boston) can be taken into account with a shift of about three hours so that
Miller's interval 3:00 a.m.$-$9:00 a.m. is made to correspond
to the range 6:00 a.m.$-$12:00 a.m. of
Jaseja et {\it al.}. If this is done, 
although one does not expect an exact correspondence
due to the difference between the 
two epochs of the year, the two characteristic trends are surprisingly
close. 

Thus we shall try to use the Miller's data 
for a rough evaluation of the ratio reported in Eq.(\ref{deltav}).
In this case, rescaling from $v_{\rm obs}$ 
to $v_{\rm earth}$ through Eq.(\ref{vobs0}) 
(for the Miller's interferometer
that was operating in air), we obtain values of
${{\delta v^2}\over{v^2}}$ in the range $0.1-0.2$. 
 This estimate, when replaced in Eq.(\ref{prediction2}) leads 
to values of $\delta(\Delta\nu)_{\rm theor}$ in the range $1.6-3.2$ kHz, 
well consistent with the value $1.6 \pm 1.2$ kHz given in
Eq.(\ref{resultb}). Of course, 
one needs more precise data. However, in spite of our 
crude approximations, the order of magnitude of the effect is correctly
reproduced. 

This suggests, once more \cite{pagano}, 
to perform a new class of ether-drift experiments in 
dielectric gaseous media. For instance,
using stabilizing cavities as in Refs.\cite{brillet,muller}, one could
replace the high vacuum 
in the Fabry-Perot with air. In this case, where
$|B_{\rm vacuum}|\sim 4 \cdot 10^{-9}$ would be replaced by
$|B_{\rm air}|\sim 9\cdot 10^{-4}$, there should be an increase by
five orders of magnitude in the typical value of $\Delta \nu$ with respect to
Refs.\cite{brillet,muller}.

\vskip 15 pt
{\bf 6.}~In this Letter we have re-considered
the possible existence of a preferred reference 
frame through an analysis of both classical and
 modern ether-drift experiments. The small 
observed velocities $v_{\rm obs}\sim 
8.5 \pm 1.5$ km/s for the Michelson-Morley, 
Morley-Miller and Miller experiments, 
$v_{\rm obs}\sim 3.1 \pm 1.0$ km/s for the
Illingworth experiment, and
$v_{\rm obs}\sim 1$ km/s for the Joos experiment, when corrected for the effect
of the refractive index, appear to point consistently to
a rather large kinematical Earth's velocity
$v_{\rm earth}\sim 204\pm 36$ km/s (in the plane of the interferometer). 

Therefore, 
it becomes natural to explore the existence of a preferred frame and
formulate definite predictions for the relative frequency shift $\Delta\nu$
which is measured in
the present-day experiments with cavity-stabilized lasers, upon local rotation
of the apparatus or under the Earth's rotation. In this case,
our basic relation is
\BE
{{\Delta\nu}\over{\nu}} \sim |B_{\rm medium}| {{v^2_{\rm earth}}\over{c^2}}
\EE
where 
$B_{\rm medium}\sim -3 ({\cal N}_{\rm medium}-1) $, 
${\cal N}_{\rm medium}$ being the refractive index of the gaseous dielectric
medium that fills the cavities. For a very high vacuum, using the 
prediction of General Relativity
   $|B_{\rm vacuum}| \sim 4 \cdot 10^{-9}$, and the range
of kinematical Earth's velocity $v_{\rm earth}\sim 204\pm 36$ km/s
suggested by the classical ether-drift experiments, we predict
${{\Delta\nu}\over{\nu}} \sim (1.9 \pm 0.7)\cdot 10^{-15}$, consistently
with the experimental result of Ref.\cite{muller}.  
For He-Ne masers, the same range of Earth's velocities leads to predict
a typical value
$\Delta\nu\sim 16$ kHz, for which
${{\Delta\nu}\over{\nu}} \sim 5\cdot 10^{-11}$,  with a characteristic
modulation of a few kHz 
in the period of the year and for the hours of the day when
Jaseja et al.\cite{jaseja} performed their experiment. This prediction is
consistent with their data, although the rather large experimental 
errors require further experimental checks. For this reason, we propose to 
replace the high vacuum adopted
in the Fabry-Perot cavities with air. In this case, where the anisotropy 
parameter $|B_{\rm vacuum}|\sim 4 \cdot 10^{-9}$ would be 
replaced by $|B_{\rm air}|\sim 9\cdot 10^{-4}$, there should be an increase by
{\it five orders of magnitude} in the typical 
value of $\Delta \nu$ with respect to
Refs.\cite{brillet,muller}. If this is not observed, the 
existence of a preferred frame will be definitely ruled out.

\vskip 25 pt
\centerline{\bf{Acknowledgements}}
We thank L. Pappalardo for many useful discussions.
\vfill
\eject

\end{document}